\documentclass[%
 amsmath,amssymb,twocolumn,
reprint,%
]{revtex4-1}

\usepackage{color}
\usepackage{graphicx}
\usepackage{dcolumn}
\usepackage{bm}

\begin{document}


\title[]
{Decoupling of translational and rotational diffusion in quasi-2D
colloidal fluids}

\author{Skanda Vivek}
\email{skanda.vivek@gmail.com}
\affiliation{Department of Physics, Georgia Institute of Technology, Atlanta, Georgia 30332
}%
\author{Eric R. Weeks}%
\affiliation{%
Department of Physics, Emory University, Atlanta, GA 30322
}%
 

\date{\today}

\begin{abstract}
We observe the translational and rotational diffusion of dimer
tracer particles in quasi-2D colloidal samples.  The dimers are
in dense samples of two different sizes of spherical colloidal
particles, with the area fraction $\phi$ of the particles varying
from dilute to nearly glassy.  At low $\phi$ rotational and
translational diffusion have a ratio set by the dimer size, as
expected.  At higher $\phi$ dimers become caged by their neighboring
particles, and both rotational and translational diffusion slow.
For short dimers we observe rapid reorientations so that the
rotational diffusion is faster than translational diffusion:  the
two modes of diffusion are decoupled and have different $\phi$
dependence.  Longer dimers do not exhibit fast rotations, and
we find their translational and rotational diffusion stay coupled for
all $\phi$.    Our results bridge prior results that used spheres
(very fast rotation) and long ellipsoids (very slow rotation).
\end{abstract}

\maketitle

\section{\label{sec:intro}Introduction}
A comprehensive explanation for the dramatic increase
in viscosity on approaching the glass transition
is still lacking, although there are a variety of
theories.\cite{biroli13,ediger12,cavagna09}  What makes the
matter more complicated is the inadequacy of the traditional
concept of viscosity and its relation with microscopic diffusion in
supercooled liquids.  In liquids, the Stokes-Einstein-Sutherland
equation\cite{einstein1905,sutherland1905} relates microscopic
translational diffusion ($D_T$) as inversely proportional to
solvent viscosity $\eta$.  Rotational diffusion ($D_R$) is also
inversely proportional to $\eta$, known as the Stokes-Einstein-Debye
relation.  Further, the ratio of translational to rotational
diffusion constants ($D_T/D_R$) should be independent of
viscosity and temperature.  However, many experiments and
simulations\cite{fujara1992,chang1994,berthier2004,kumar2006} have
shown a violation of the Stokes-Einstein and Stokes-Einstein-Debye
relations in supercooled liquids.  These violations can manifest
as $D_T$ and/or $D_R$ no longer being inversely related to
$\eta$, and also $D_T/D_R$ no longer being a constant.

Pioneering experiments in the early 1990's observed a violation
of these relations on approaching the glass transition in
orthoterphenyl.\cite{fujara1992,
chang1994}  Rotation and translational diffusion constants were
measured indirectly through spin-relaxations.
They observed an enhancement of
translation relative to rotation approaching the glass transition.
At the time, this strange difference was attributed to spatial
distribution of relaxation timescales $\rho(\tau)$, and measured
rotation and translation measurements being sensitive to different
moments of this distribution.\cite{ediger2000}  But it was
thought that on the single molecule scale, translation and rotation
remain coupled.

However, recent simulations and colloidal experiments have
found that decoupling occurs even at the single particle
level.\cite{chong2005,chong2009,kim2011,edmond2012}
The current hypothesis is that decoupling occurs due to
translation and rotation degrees encountering different
dynamic length scales.\cite{ediger2012}  Moreover, some
studies found that translation was enhanced relative
to rotation,\cite{zheng11,edmond2012,zheng2014}
whereas others found rotation was enhanced relative to
translation.\cite{chong2005,chong2009,kim2011}  These different
experiments had different probe shapes and conditions so direct
comparison of the observations is challenging.

In this study we examine how the probe details influence
translational-rotational decoupling in colloidal samples.
Colloidal samples at high concentration have been established
as model glass formers,\cite{Pusey86,weeks17} and have the
advantage that individual particles can be visualized.  Here we
use naturally occurring anisotropic silica dimers of different
aspect ratio as tracers, and find that the dimer length determines
translation-rotation decoupling.

\begin{figure}
\centering
\includegraphics[width=0.6\columnwidth]{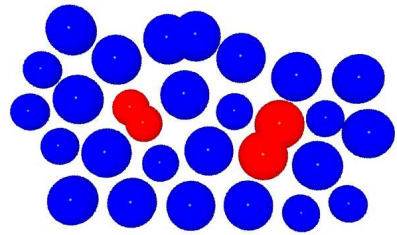}
\caption
{Example of a short dimer ($l_l=3.92 \mu$m) and long dimer
($l_l=5.37 \mu$m) in a concentrated sample with area fraction
$\phi=0.79$. The rendering is adapted from particle positions. Dimers
are in red, and neighbors in blue.  We find short dimers
can rotate more easily, whereas long dimers are more constrained
by their neighbors.
}
\label{fig:img2in1}
\end{figure}

These dimers are at a very low concentration, less than 5 percent
and the experiments are the same as previously published (where
we studied only the monomer particles\cite{vivek17}).  We find
that for dimers of smaller lengths, rotations do not slow down
as much as translations, on approaching the glass transition,
similar to the case of spheres as probes.\cite{kim2011}
However we find that in our longest dimers, $D_T$ and $D_R$
remain coupled at all concentrations, i.e. $D_T/D_R\sim$ constant.
Figure \ref{fig:img2in1} shows an example computer rendering of a
long dimer and short dimer in one of our samples.  Our key result
is that the shorter dimer can rotate more easily as it is easier
to relax the steric hindrance of the neighboring particles.

\section{\label{sec:mat}Materials and methods}
\subsection{\label{sec:sample}Experimental methods}

For this work we reanalyze movies corresponding to previously
published data.\cite{vivek17}  In the experiments we use gravity
to confine bidisperse non-functionalized silica particles to a
monolayer (diameters $\sigma_S=2.53$ and $\sigma_L=3.38$ $\mu$m,
Bangs Laboratories, SS05N). The number ratio is $N_L/N_S = 1.3 \pm
0.5$, and varies from sample to sample.  The control parameter is
the area fraction $\phi$, with glassy samples found for $\phi
> 0.79$; the data we present here are all with $\phi < 0.79$.
The particles are sedimented to the microscope coverslip of our
sample chamber prior to observation.  The coverslip is made
hydrophobic by treatment with Alfa Aesar Glassclad 18 to prevent
particle adhesion, and indeed we do not observe any particles
stuck to the glass.  We do not add salt.
We verify that in all experiments, only one layer of particles is
present (ensured by keeping the overall particle concentration below
the level that requires a second layer to form).  We use brightfield
microscopy and a CCD camera to record movies of particles diffusing.

The samples have naturally occurring dimers at dilute concentrations
from $2-5$\%.  The dimers are stable during our observations.
All dimers are made of identical particles (either two
small particles or two large particles), indicating they are
formed prior to the experiment; they seem to be present in the
samples as received.  Rather than being two spheres barely
touching, they are somewhat fused together, as can be seen in 
Fig.~\ref{fig:dimers-4in1}$A$.  The aspect ratio (length/width)
is always less than two, and varies from dimer to dimer.

\subsection{\label{subsec:dimers}Imaging and tracking dimers}

\begin{figure}
\centering
\includegraphics[scale=.3]{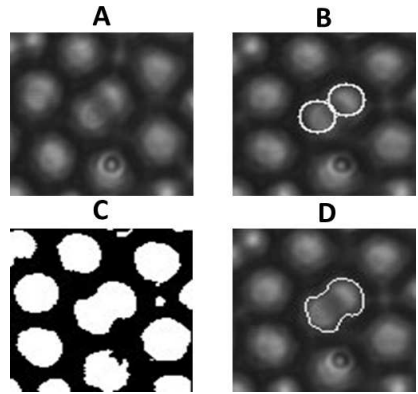}
\caption
{($A$) Image of a dimer surrounded by spherical particles. ($B$)
Identification of two individual particles in a dimer.  ($C$)
Thresholded image. ($D$) Outline of dimer from the thresholded
image.  The aspect ratio of the dimer is 1.65, and the size of
the image is $\sim 10\mu$m.}
\label{fig:dimers-4in1}
\end{figure}

We need to follow the translational and rotational motion
of the dimers.  We start by using standard particle tracking
software\cite{idlref} to track the two particles of a dimer,
as shown in Fig.~\ref{fig:dimers-4in1}$B$.  Next we select
the region of the image (ROI) that included dimers.  Based on
brightness, we threshold this ROI to get a black and white image
(Fig.~\ref{fig:dimers-4in1}$C$).  We identify connected regions in
this thresholded image, and selected the largest such region as the
dimer of interest.  The length of the longest axis of this connected
region is measured as $l_l$, and then we identify the length of the
short axis $l_s$ as the longest distance across the connected region
perpendicular to the long axis.  The aspect ratio then is $l_l/l_s$.

Recent work has shown that two-dimensional glass-forming systems
need to be analyzed slightly differently than 3D 
systems.\cite{vivek17, shiba16, illing17}  The key concern is that
2D systems are subject to Mermin-Wagner fluctuations which move
particles locally but do not lead to structural 
rearrangements.\cite{peierls34,landau37,mermin66,mermin68,klix15,illing17}
Particles are ``caged'' by their neighbors, but Mermin-Wagner
fluctuations result in coherent motion of the cage.
Analyzing the motion of particles relative to their caging neighbors
removes the influence of these fluctuations, making apparent the
motions that lead to structural relaxation.\cite{MaretEPL2009}

Accordingly, to determine the relaxation time scale for our samples,
we define the cage relative translational correlation function as:
$F_{S-CR}(k,\Delta
t)=\langle\exp(i\vec{k}\cdot \Delta\vec{r}_{CR})\rangle_t$ where
$\Delta\vec{r}_{CR}=\vec{r}(t+\Delta
t)-\vec{r}(t)-\frac{1}{N}\sum_j[\vec{r_j}(t+\Delta
t)-\vec{r_j}(t)],$
where $j$ denotes the nearest neighbors of the particle at the
initial time $t$, and the sum is over all neighbors.  The $\alpha$
relaxation timescale $\tau_{\alpha}$ is the timescale when
$F_{S-CR}$ reaches $1/e=0.37$, and defines the time scale over which
the sample has significant structural rearrangements.\cite{vivek17}

The cage relative mean square displacement (MSD) is defined using the
same displacements $\Delta\vec{r}_{CR}$. We measure the long time
translational diffusion coefficient $D_T$ from the cage relative
MSD. For our tracers, we do not see a significant difference
between the MSD and the cage-relative MSD in the observed area
fraction range. However, softer samples are known to have larger
differences.\cite{vivek17, illing17}  We are interested in the
$\Delta t \rightarrow \infty$ behavior, so distinctions between
motion along the dimer axis and perpendicular to that axis will
not be important to us.\cite{han06}  Rotational mean square
displacements (MSD$_{\rm R}$) do not require cage-relative
analysis.  For the MSD$_{\rm R}$ we identify the instantaneous
angle $\theta(t)$ of a dimer (in radians); unwrap this angle so
that it can take values smaller than 0 or larger than $2\pi$;
and then compute the MSD$_{\rm R}$ from this unwrapped angle.

\subsection{\label{subsec:theory}Hydrodynamic theory}

The dimers we analyze have an aspect ratio $l_l/l_s < 2$,
whereas perfect dumbbells have aspect ratio 2.  Nonetheless,
a reasonable starting approximation is to model our dimers as
dumbbells.  For dumbbells in a liquid the ratio of translational to
rotational diffusion coefficients is given as\cite{cugliandolo2015} 
\begin{equation}
\frac{D_T}{D_R}=\frac{\sigma^2}{4}
\label{eq:dimerdiff}
\end{equation}
\noindent
where $\sigma$ is the diameter of particles in the dumbbell.
This predicts a ratio of 1.60~$\mu$m$^2$ for a dimer composed of
two small particles in our experiment and 2.86~$\mu$m$^2$ for a
dimer composed of two large particles.

\begin{figure}
\centering
\includegraphics[width=0.8\columnwidth]{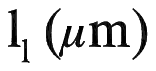}
\caption
{$D_T/D_R$ as a function of the dimer length $l_l$, from a medium-concentration
sample with $\phi=0.55$ 
($\tau_{\alpha}=9.5$~s).
The dashed line is Eqn.~\ref{eq:dimerdiff} using $\sigma = l_l / 2$.
Note that for these data, we are at low area fraction and so we use
the normal MSD rather than the cage-relative MSD to measure
translational diffusion constants.
}
\label{fig:Dcr-l-tau0}
\end{figure}

Figure~\ref{fig:Dcr-l-tau0} shows $D_T/D_R$ as a function of
dimer length $l_l$ in a medium-concentration sample.  The data are scattered
with no systematic dependence on $l_l$.  The dashed line shows
Eqn.~\ref{eq:dimerdiff} and the data are within a factor of
two of this prediction.  This is reasonable, as (1) our dimers
are not the perfect dumbbell shape as assumed by theory, and (2)
Eqn.~\ref{eq:dimerdiff} does not take into account the hydrodynamic
influence of the bottom wall,\cite{sarmientogomez16} which affects translational and
rotational modes differently.\cite{han06,bhattacharya05}  It is
also possible that the different values of $l_l$ for dimers
ostensibly made from identical particles give rise to dimers
of different shapes (different amounts of overlap of the two
particles) which could account for the scatter in our data.  In sum,
we recognize an inherent uncertainty for the ratio of $D_T/D_R$
of about a factor of 2, and will look for this ratio to vary by
more than a factor of two as we approach the glass transition.

\section{\label{sec:results}Results}

\begin{figure}
\centering
\includegraphics[scale=.35]{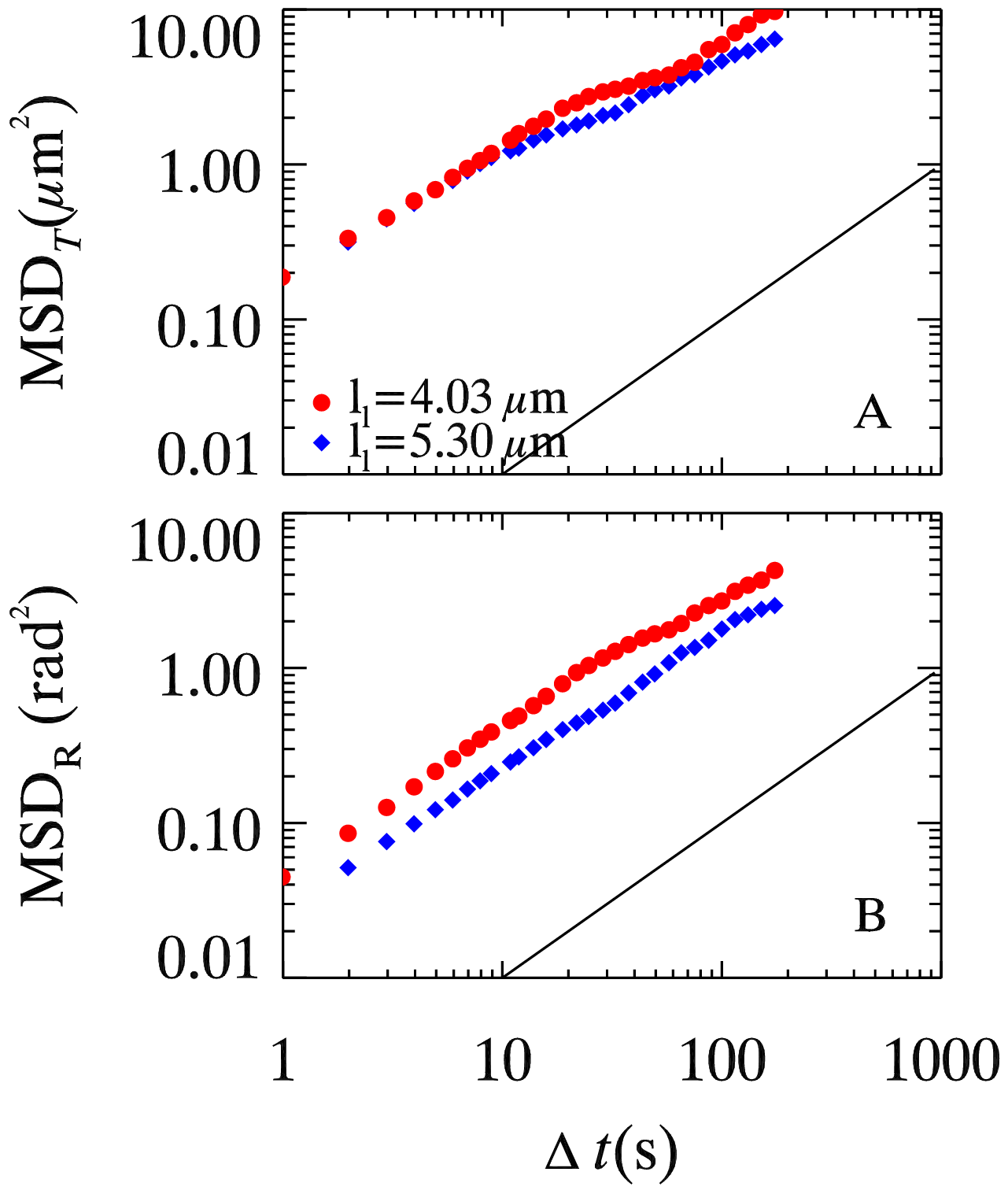}
\includegraphics[scale=.35]{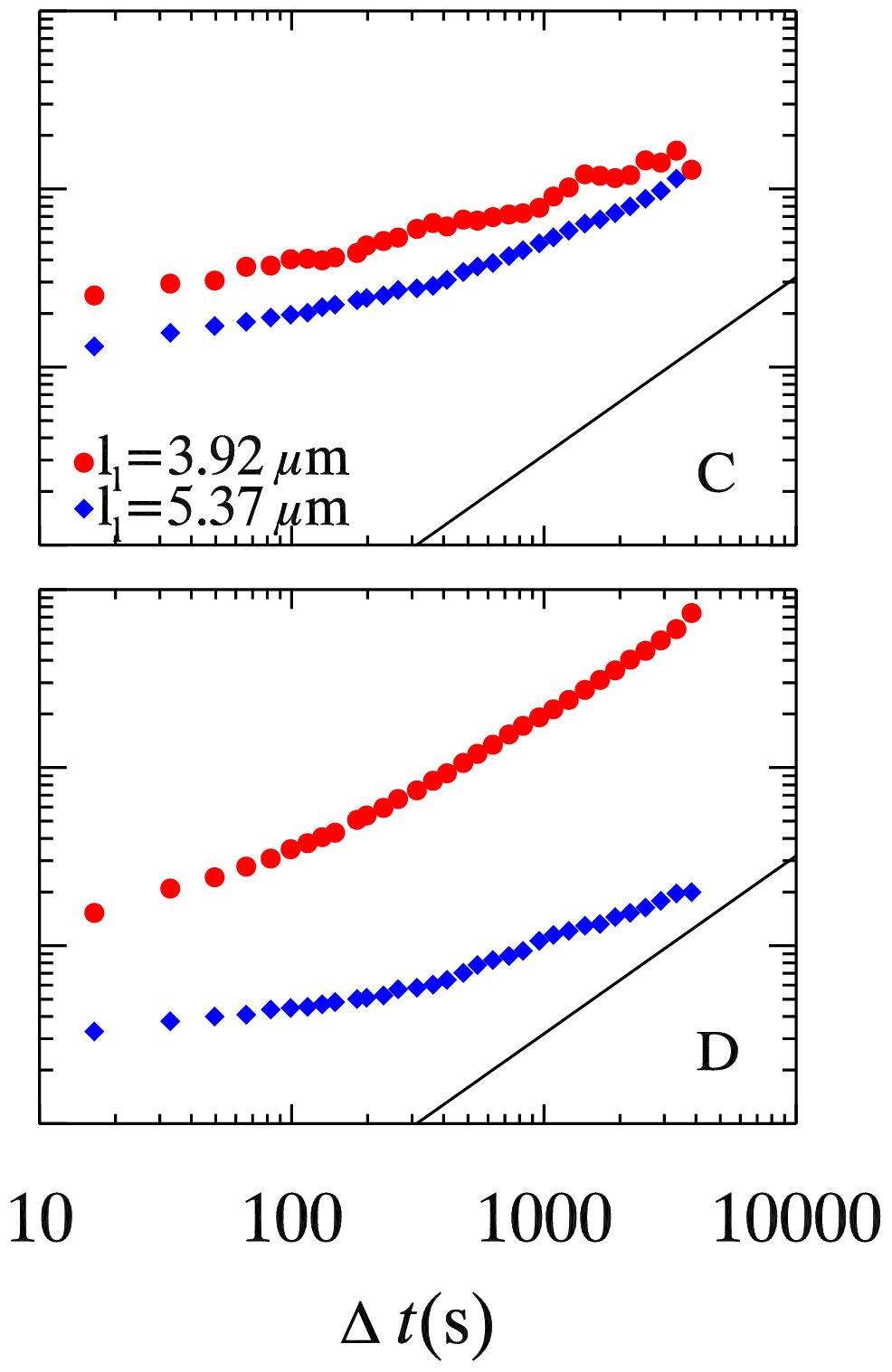}
\caption
{($A$) Cage-relative translational MSD of a short ($l_l=4.03$
$\mu$m, aspect ratio=1.76) and long ($l_l=5.30$ $\mu$m, aspect ratio=2.00) dimer in a slightly
supercooled sample at $\phi=0.55$. $D_T=0.015$~$\mu$m$^2$/s and
$0.009$~$\mu$m$^2$/s for the short and long dimers respectively.
($B$) Rotational MSD of the same dimers as in ($A$).
$D_R=0.017$~1/s and 0.01~1/s for the short and long dimers
respectively.
($C$) Cage-relative translational MSD of a short ($l_l=3.92$
$\mu$m, aspect ratio=1.96) and long ($l_l=5.37$ $\mu$m, aspect ratio=1.80) dimer in concentrated sample
at $\phi=0.79$. $D_T=0.0015$~$\mu$m$^2$/s and $0.00008$ $\mu$m$^2$/s
for the short and long dimers respectively.
($D$) Rotational MSD of the same dimers as in ($C$).
$D_R=0.001$~1/s and 0.00003~1/s for the short and long dimers
respectively.  In all panels, the black lines indicate a slope of 1,
which is the case for diffusive behavior.
}
\label{fig:msd}
\end{figure}

Figure~\ref{fig:msd} shows translational and rotational MSDs
in medium concentration ($A$,$B$) and high concentration ($C$,$D$)
samples corresponding to different length dimers.  For the medium
concentration samples ($\phi = 0.55$) all MSDs rise diffusively,
MSD $\sim \Delta t$, as can be seen by comparing the data to the
straight black line which has slope = 1.  The long dimer diffuses a
bit slower, as seen by the blue diamonds lying below the red circles
in ($A$,$B$). Interestingly at large concentration ($\phi=0.79$), 
we see that the translational
MSD is similar for the short and long dimers (Fig.~\ref{fig:msd}$C$), while the rotational MSD is
much different (Fig.~\ref{fig:msd}$D$). Comparing Fig.~\ref{fig:msd}$A$
and $C$, we see that both long and short dimers show a similar slow
down in translation.  However, Fig.~\ref{fig:msd}$B$ and $D$ are very
different. Here, long dimers (blue diamonds in this graph) show
a much larger slowdown in rotation as compared with short dimers
(red circles) at high concentration.

\begin{figure}
\centering
\includegraphics[scale=.55]{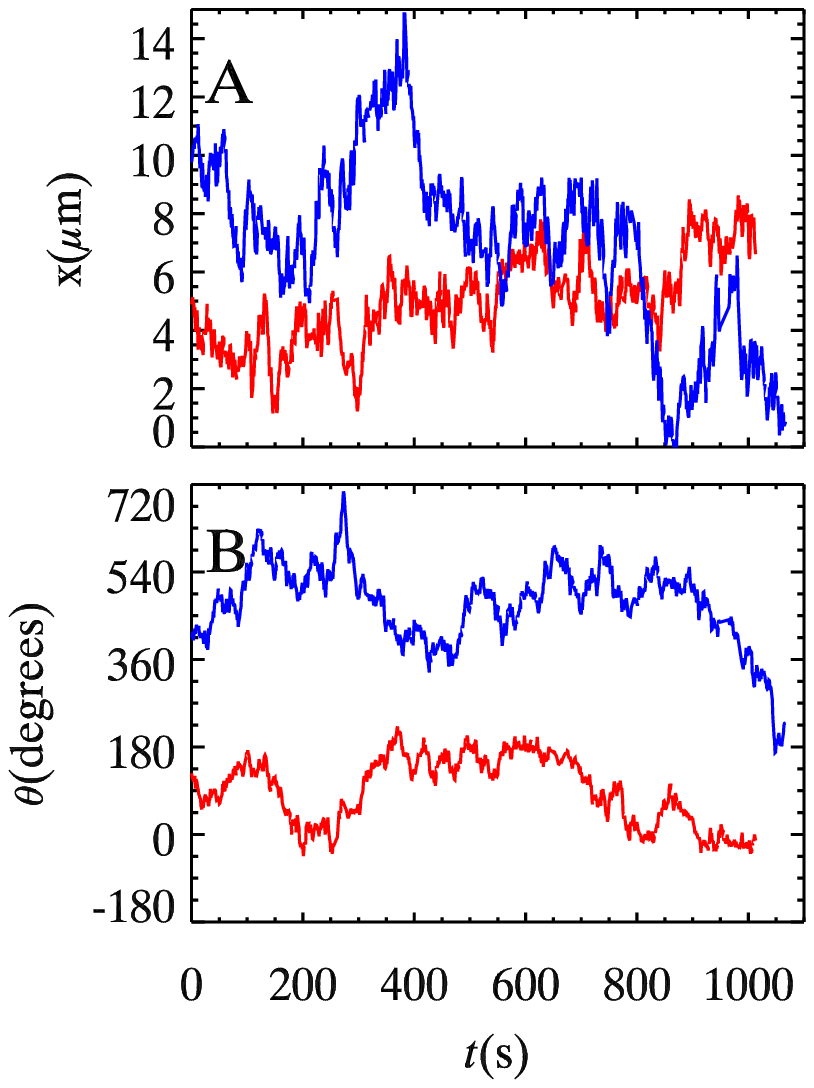}
\includegraphics[scale=.55]{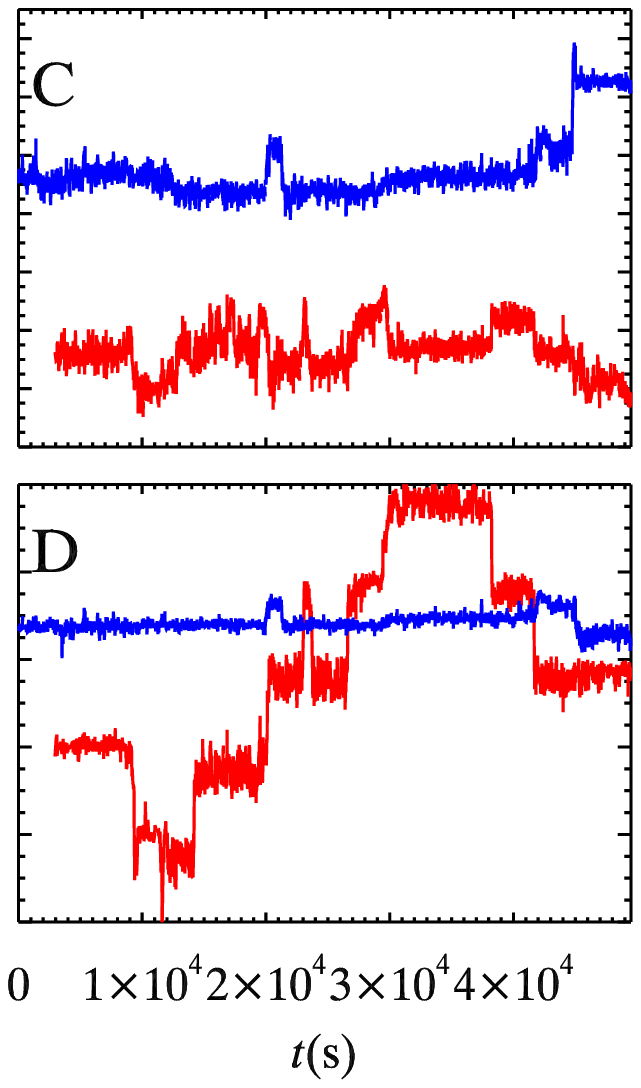}
\caption
{($A$) Cage-relative displacements in the $x$ direction of a short 
($l_l=4.03$ $\mu$m in red)
and long ($l_l=5.30$ $\mu$m in blue) dimer in a slightly
supercooled sample at $\phi=0.55$.  These are the dimers analyzed
in Fig.~\ref{fig:msd}$A$ and $B$.  
($B$) Angular trajectories corresponding to $A$.
($C$) Cage-relative displacements in $x$ of a short
($l_l=3.92$ $\mu$m in red) and long ($l_l=5.37$ $\mu$m in blue)
dimer in concentrated sample at $\phi=0.79$.  These are the dimers
analyzed in Fig.~\ref{fig:msd}$C$ and $D$.  
($D$) Angular trajectories corresponding to ($C$).
Note that all trajectories
are shifted to put each dimer onto the same graph; they are not
actually adjacent dimers.
}
\label{fig:traj}
\end{figure}

Figure~\ref{fig:traj} shows the trajectories of the dimers
corresponding to Fig.~\ref{fig:msd}.  For the medium concentration
sample ($A$), long and short dimers have similar trajectories.
In contrast, the concentrated sample ($C$) shows that the long
dimer (blue) spends more time localized (perhaps with a
reversible cage rearrangement event in the
middle\cite{vollmayrlee04} and an irreversible jump at the end).
The shorter dimer (red) moves more frequently.

Additionally, we track the angular motion of each dimer, shown
in Fig.~\ref{fig:traj}$B$ and $D$.  This is easy to measure as we
can track both particles in the dimer, and hence compute the
instantaneous angle $\theta$.
We keep track of the number of rotations so that $\theta$ is
unbounded.  In the medium concentration sample,
shown in Fig.~\ref{fig:traj}$B$, long and short dimers have similar
angular displacements.  However in the concentrated sample shown
in Fig.~\ref{fig:traj}$D$, clearly the short dimer (red) shows much
larger angular displacements than the long dimer (blue). Thus the
small dimer is able to rotate much easier than the long dimer.
In particular, it often rotates by $180^\circ$, suggesting that
the cage of neighboring particles expands slightly, the dimer has
a chance to rotate, and then the cage contracts locking the dimer
into the original orientation or else $180^\circ$ rotated.

\begin{figure}
\centering
\includegraphics[width=\columnwidth]{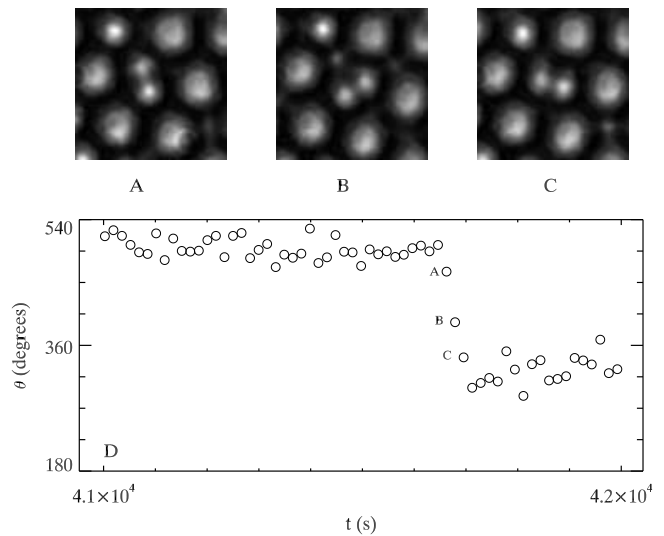}
\caption
{Angular displacements of short dimer, same as shown in~\ref{fig:traj}$D$.
Images show corresponding dimer during certain short time
interval, from ($A$) to ($C$).  The size of
each image is $\sim 10\mu$m.
($D$) shows the zoomed in interval of large angular displacement,
where $A,B,$ and $C$ are the same time points
as the respective images.
}
\label{fig:theta}
\end{figure}

The short dimer shows jumps in angle at certain times.  
Figure \ref{fig:theta} shows this dimer at
three consecutive time points $A$, $B$, and $C$ during which a
large jump takes place. The time between each is the recording
rate of 16.5 seconds per image (used for the concentrated sample).
$D$ shows this characteristic jump zoomed in. Clearly, the dimer
rotates fairly quickly. While it is possible we miss rotations that
are more than $180^\circ$ between video frames, we believe this
is unlikely.  We examine the probability distributions of angular
jumps between video frames, $P(|\Delta \theta|)$, and find that the
probability of large jumps is nearly zero for $|\Delta \theta| >
170^\circ$.  This suggests that larger jumps with $|\Delta \theta|
> 180^\circ$ are even rarer.

\begin{figure}
\centering
\includegraphics[width=0.9\columnwidth]{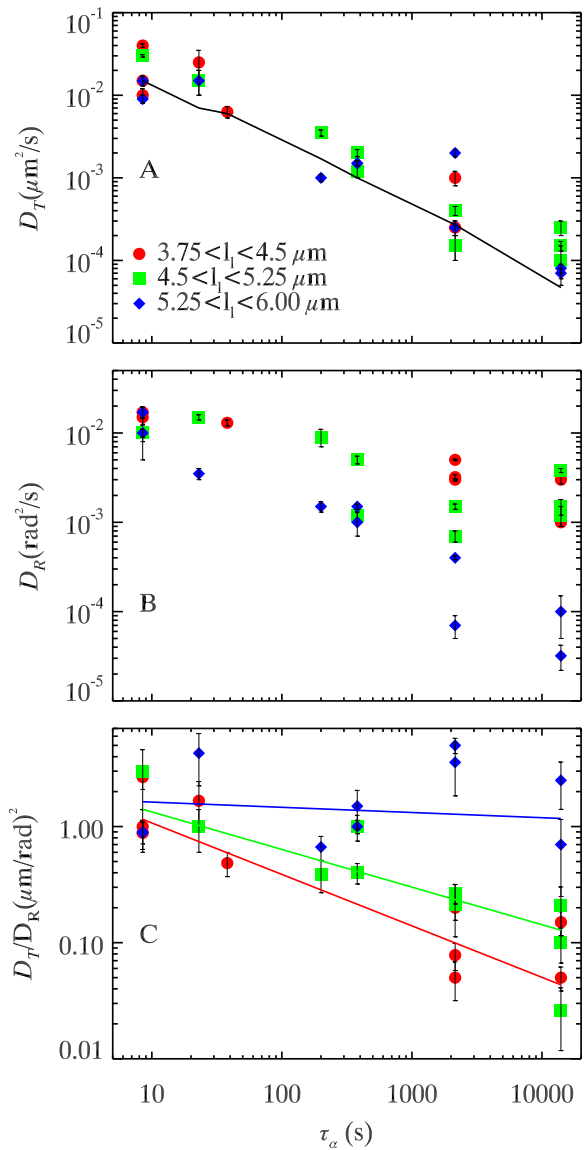}
\caption
{Diffusion coefficients as a function of $\tau_{\alpha}$
for different dimer lengths.  ($A$) Translational diffusion
constants $D_{T}$ . The black line denotes the measured bulk
sample $D_T$, that is,
the diffusion coefficient of the spherical bath particles.  ($B$)
Rotational diffusion constants $D_R$.  ($C$) The ratio $D_{T}/D_R$.
In all panels, the different colors denote different ranges of
$l_l$ as indicated in the legend in panel ($A$).  The
lines in $C$ are power-law fits.
}
\label{fig:Dcr-ta-longaxis}
\end{figure}

To compare all our data we fit the large $\Delta t$ behavior of
all MSD curves to measure diffusion constants.  The translation
diffusion coefficient $D_T$ is measured as $\langle \Delta
r_{CR}^2(\Delta t) \rangle =4D_T \Delta t$ (using cage-relative
displacements).  The average is over all particles and all
initial times.  The rotational diffusion coefficients are
measured similarly, through $\langle \Delta \theta^2(\Delta t) \rangle
=2D_R \Delta t$, where $\Delta \theta$ is angular displacement
in $\Delta t$.  Figure~\ref{fig:Dcr-ta-longaxis}$A$ plots $D_T$
as a function of $\tau_{\alpha}$ (defined in Sec.
\ref{sec:mat})
for all samples. Here we see that all dimers follow the black line,
which is the measured bulk $D_T$ of the spherical particles in
the sample.  A fit to the bulk data (not shown) finds $D_T \sim
\tau_{\alpha}^{-0.73}$, which is in the range of exponents observed
in other experiments.\cite{mazza2007,chong2009}

Rotational diffusion also slows for glassy samples as shown by the
data for $D_R$ in Fig.~\ref{fig:Dcr-ta-longaxis}$B$.  In contrast
with $D_T$, these data become more scattered for the samples
closer to the glass transition (larger $\tau_\alpha$).  
It is apparent that the slowest rotational diffusion is seen for
the longest dimers (blue diamonds) which slow by $\sim 10^2 - 10^3$
as $\tau_\alpha$ grows by $10^3$.  In contrast, the shortest dimers
(red circles) decrease by only $\sim 10^1$ over the same range.
The difference between $D_R$ of the long and short dimers is
more than an order of magnitude at the largest $\tau_{\alpha}$.
The mild decrease in $D_R$ for the short dimers is similar to that
seen with spherical colloids in a prior experiment.\cite{kim2011}
Overall, this is what we expect based on the conceptual
sketch of Fig.~\ref{fig:img2in1}:  long dimers require their
neighbors to move out of the way to facilitate their rotation,
whereas short dimers are constrained less by their neighbors.
Likewise this is supported by Fig.~\ref{fig:traj}$D$ where the
short dimer makes large jumps in angle.

We examine decoupling of rotational and translational diffusion 
by plotting $D_T/D_R$ in Fig.~\ref{fig:Dcr-ta-longaxis}$C$.
The long dimers (blue diamonds) show a constant $D_T/D_R$
independent of $\tau_{\alpha}$; here the two diffusion constants
are coupled at all area fractions.  In contrast, the short dimers (red circles)
show a decrease in $D_T/D_R$ with increasing $\tau_{\alpha}$.  The colored
lines show power-law fits for the three different dimer length
regimes.  
Some of the observed spread in data could be due to dynamical heterogeneity.
Closer to the glass transition, 
dynamical heterogeneity causes differences in diffusivities across
the sample.\cite{weeks_science} This could cause different particles
of similar shape to differ in $D_T/D_R$ depending on their
local environment.  However, at the largest $\tau_{\alpha}$,
the difference between $D_T/D_R$ of the long and short dimers is
more than an order of magnitude.  We see decoupling:  while both
rotational and translational diffusion slow as the glass transition
is approached, rotational diffusion slows less dramatically --
at least for the short dimers.

\begin{figure}
\centering
\includegraphics[width=0.9\columnwidth]{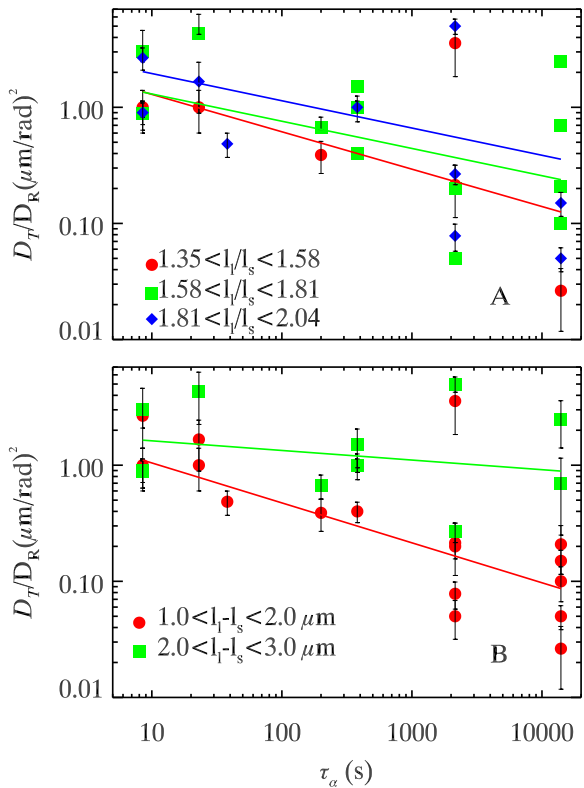}
\caption
{($A$) $D_T/D_R$ as a function of  $\tau_{\alpha}$ for 
different aspect ratio $l_l/l_s$ as indicated in the legend. 
($B$) The same data for different values of $l_l - l_s$ as
indicated in the legend. In both panels the lines are fits
to the data.
}
\label{fig:Dcr-l-fixedta2}
\end{figure}

Organizing the data based on dimer length shows a clear
trend from long dimers (no decoupling) to short dimers
(decoupling); another possible variable is the aspect ratio.
Figure~\ref{fig:Dcr-l-fixedta2}$A$ shows $D_T/D_R$ plotted as
a function of $\tau_{\alpha}$,
equivalent to Fig.~\ref{fig:Dcr-ta-longaxis}$C$, but here
different colors denote aspect ratio instead of dimer length.
There is no clear trend with aspect ratio, in strong contrast to
Fig.~\ref{fig:Dcr-ta-longaxis}$C$.  Hence the longest axis seems to
be the relevant parameter, rather than aspect ratio.  A final way to
think about the data is motivated by Fig.~\ref{fig:traj}$D$ showing
that the short dimers can rotate by $180^\circ$; this is presumably
because their cage of neighboring particles expands slightly,
allow the rotation.  The expansion distance can be estimated as
$l_l - l_s$, arguing that the neighbors start $\sim l_s$ away from
the middle of the dimer and expand to $\sim l_l$ to allow the long
axis to rotate past them.  This suggests that the $l_l$ dependence
[Fig.~\ref{fig:Dcr-ta-longaxis}$C$] could be a dependence on $l_l
- l_s$.  This seems plausible; Fig.~\ref{fig:Dcr-l-fixedta2}$B$
shows the data with color indicating different ranges of $l_l
- l_s$, which reasonably well separates the faster rotating
dimers (red dimers, small $l_l-l_s$) from the slower rotating
dimers (green squares, large $l_l - l_s$).  The fact that $l_l$
works slightly better [Fig.~\ref{fig:Dcr-ta-longaxis}$C$] may be
because the number of neighbors that need to move scales as $l_l$,
independent of $l_s$.  Note that the slight cage expansion
allowing the $180^\circ$ rotation is likely unrelated to
dynamical heterogeneity.  While the cage expansions are uncommon
events, the fluctuations in cage size occur on a short time scale
which is not directly related to long-time-scale rearrangement
motions.\cite{weeks_science}

Previous simulations found that quantifying
translation-rotation decoupling depends on the analysis
method.\cite{chong2009,lombardo2006}  The ``Debye method''
uses dot products of the initial and final orientation of a
tracer; using this method to measure $D_R$ changed the nature of
decoupling.\cite{chong2009,lombardo2006}  We also measured $D_R$
from this formalism, in the same way as by Edmond~\textit{et
al}.\cite{edmond2012}  We find that both methods to measure
$D_R$ give the same result within error, as also found by
Edmond~\textit{et al}.

Our results indicate that the longest dimension controls how
the particle rotation is sterically constrained by the neighbors
on approaching the glass transition.  Figure \ref{fig:img2in1}
suggests this may be because the longest dimension determines
how many neighboring particles can restrict rotational motion.
Of course, all of our results are for a particular colloidal
sample with particles having a size ratio $1:1.34$.  It is
possible that the results would differ in other samples.  With a
larger size ratio, cages surrounding dimer tracers would vary
strongly in composition which might change the frequency of the
fluctuations that allow $180^\circ$ rotations such as
Fig.~\ref{fig:theta}.\cite{KuritaPRE2010}

\section{\label{sec:disc}Discussion}

Prior groups have studied glass transitions in colloidal glasses
composed of anisotropic particles.\cite{zheng11,zheng2014,kim2011}
Kim {\it et al.}\cite{kim2011} studied rotation and translation
of optically anisotropic spheres (aspect ratio 1).  Here, $D_T/D_R$
in the concentrated regime was almost 2 orders of magnitude smaller
than the dilute limit.\cite{kim2011} As spheres rotate without
any steric hindrance from neighboring spheres, in this experiment,
translation slows down more than rotation approaching the glass
transition.\cite{weeks17,kim2011}  More precisely, translation
slow down dramatically, but rotation only slows slightly.

Other experiments by Zheng {\it et al.} looked at ellipsoids of
aspect ratios varying from 2.3 to 6.\cite{zheng11,zheng2014}
For the small aspect ratio of 2.3, $D_T/D_R$ did not change
approaching the glass transition, similar to what we see with our
long dimers.  For the large aspect ratio 6, however, $D_T/D_R$
was an order of magnitude higher than the dilute limit, indicating that
rotation slows down more than translation.  This study concluded
that the decoupling with enhanced translational motion occurs
for situations with aspect ratio $\gtrsim 2.5$.\cite{zheng2014}
A similar observation of enhanced diffusion was seen by Edmond {\it
et al.} who studied tetrahedral cluster tracers in a 3D sample of
spherical particles.\cite{edmond2012}  There, the ratio between
the longest dimension of the cluster to the mean particle size
was 2.9.  While a direct comparison between 2D ellipsoids and 3D
tetrahedra seems dubious, nonetheless the observations of Edmond
{\it et al.}\cite{edmond2012} is in conceptual agreement with the
observation of Zheng {\it et al.}\cite{zheng2014}

Our dimer experiments bridge the gap between aspect ratio
$\sim 1.3-2$.  We see that close to the glass
transition, small changes in dimer length cause significant
changes in rotational diffusion.  Smaller dimers show a weaker
slowdown in rotation on approaching the glass transition.  This is
because smaller dimers rotate more freely even in a dense sample.
Figure \ref{fig:traj}$D$ shows that this easier rotation is likely
due to slight motions of neighboring particles which momentarily
allow a rotation of $180^\circ$ for the short dimers. The
results of Kim {\it et al.}\cite{kim2011} with spheres are the
logical limit of our results, where a particle can rotate freely
with only hydrodynamic interactions with neighbors, but no steric
hindrance to rotation. Summarizing all of the experimental
observations from short to long particle shape:
there are spheres that rotate easily,\cite{kim2011} 
short dimers that can make rapid jumps
(Fig.~\ref{fig:theta}), longer dimers with coupled translational
and rotational diffusion (blue diamonds of
Fig.~\ref{fig:Dcr-ta-longaxis}$C$),\cite{zheng11,zheng2014}, and
still longer particles for which rotations are strongly inhibited
and slower than translational motion.\cite{edmond2012,zheng11,zheng2014}
Increasing the particle length relative to the cage size 
changes from decoupled fast rotation to coupled
translation/rotation to decoupled slow rotation.  Of course,
length is only one aspect of particle anisotropy;\cite{glotzer07}
our results suggest that steric hindrance from the cage
surrounding a tracer is a useful idea which may inform coupling
or decoupling of more complex particles.

In our experiments, translational diffusion is not 
affected as much by the dimer length.  This is in marked
contrast to previous experiments in polymer glasses, where
translational diffusion was found to be affected by
tracer shape, and
not rotational diffusion;\cite{hall1998} in polymer experiments,
translational diffusion slowed down more than rotational
diffusion as the glass transition was approached.  These results
are different from what we see, but more like the long ellipsoid
experiments.\cite{zheng11,zheng2014}

In summary, our results bridge the prior colloidal observations,
and collectively these observations show that steric interactions
affecting rotational diffusion depend in an important way on the
longest dimension of the tracer particles.  This highlights the
importance of steric interactions for understanding decoupling of
translational and rotational diffusion near the glass transition.

We thank J.~C.~Crocker for helpful discussions.  
This work was supported by the National Science Foundation
(CMMI-1250235 for S.V. and DMR-1609763 for E.R.W.).

\end{document}